\documentclass{WileyMSP-template}
\usepackage{graphicx}
\usepackage{amssymb}
\usepackage[super,square,comma]{natbib}
\graphicspath{{Figures/}}
\usepackage{textcomp}
\usepackage{siunitx}
\usepackage{xcolor}
\usepackage{xspace}
\usepackage[noabbrev]{cleveref}
\usepackage[normalem]{ulem}
 \usepackage{float}

\newcommand{\AD}{$A_\text{D}$\xspace}
\newcommand{\FD}{$F_\text{D}$\xspace}
\newcommand{\FDo}{$F_\text{D}$}
\newcommand{\FMu}{$F_\text{Mu}$\xspace}
\newcommand{\FMuo}{$F_\text{Mu}$}
\newcommand{\AMu}{$A_\text{Mu}$\xspace}

\newcommand{\Muz}{Mu$^{0}$\xspace}
\newcommand{\MuT}{Mu$_{T}$\xspace}
\newcommand{\MuBC}{Mu$_{BC}$\xspace}
\newcommand{\Muzo}{Mu$^{0}$}
\newcommand{\Mup}{Mu$^{+}$\xspace}

\newcommand{\Mum}{Mu$^{-}$\xspace}

\newcommand{\mup}{$\mu^{+}$\xspace}

\newcommand{\sio}{SiO$_{2}$\xspace}
\newcommand{\sioo}{SiO$_{2}$\xspace}
\newcommand{\dit}{$D_\text{it}$\xspace}

\newcommand{\LEmuSR}{LE-$\mu$SR\xspace}
\newcommand{\muSR}{$\mu$SR\xspace}

\begin{document}

\pagestyle{fancy}
\rhead{\includegraphics[width=2.5cm]{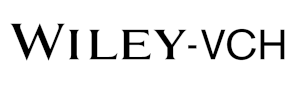}}

\title{Defect Profiling of Oxide-Semiconductor Interfaces Using Low-Energy Muons}
\maketitle
% Author: Please give full first and last names for authors and include * after the name of all corresponding authors

\author {Maria In{\^e}s Mendes Martins*}
%\email[]{maria.martins@psi.ch}
%\affiliation{Laboratory for Muon Spin Spectroscopy, Paul Scherrer Institute, Forschungsstrasse 111, 5232 Villigen PSI, Switzerland }

\author{Piyush Kumar}
%\email[]{kumar@aps.ee.ethz.ch}
%\affiliation{Advanced Power Semiconductor Laboratory, ETH Zurich, Physikstrasse 3, 8092 Zurich, Switzerland}

\author{Judith Woerle}
%\email[]{woerle@aps.ee.ethz.ch}
%\affiliation{Advanced Power Semiconductor Laboratory, ETH Zurich, Physikstrasse 3, 8092 Zurich, Switzerland}

\author{Xiaojie Ni}

\author{Ulrike Grossner}
%\affiliation{Advanced Power Semiconductor Laboratory, ETH Zurich, Physikstrasse 3, 8092 Zurich, Switzerland}

\author{Thomas Prokscha}
%\email[]{thomas.prokscha@psi.ch}
%\affiliation{Laboratory for Muon Spin Spectroscopy, Paul Scherrer Institute, Forschungsstrasse 111, 5232 Villigen PSI, Switzerland}

% Dedication

\dedication{}

% Affiliations: Please provide academic titles (Prof. or Dr.) for all authors where applicable, and include an institutional email address for all corresponding authors
\begin{affiliations}
M. I. Mendes Martins, P. Kumar, Dr. J. Woerle, Prof. Dr. U. Grossner\\
Advanced Power Semiconductor Laboratory, ETH Zurich, Physikstrasse 3, 8092 Zurich, Switzerland\\
Email Address: maria.martins@psi.ch\\

M. I. Mendes Martins, Dr. X. Ni, Dr. T. Prokscha\\
Laboratory for Muon Spin Spectroscopy, Paul Scherrer Institute, Forschungsstrasse 111, 5232 Villigen PSI, Switzerland

\end{affiliations}

% Keywords: Please provide a minimum of three and a maximum of seven keywords, separated by commas

\keywords{muon spin spectroscopy, low energy muons, interface defects, silicon, silicon carbide}

% Abstract should be written in the present tense and impersonal style (i.e., avoid we), and be at most 200 words long
\begin{abstract}
%About 5% of the article length, not more than 500 words
Muon spin rotation with low-energy muons (\LEmuSR) is a powerful nuclear method where electrical and magnetic properties of surface-near regions and thin films can be studied on a length scale of $\approx$\SI{200}{\nano\meter}. 
In this work, we show the potential of utilizing low-energy muons for a depth-resolved characterization of oxide-semiconductor interfaces, i.e. for silicon (Si) and silicon carbide (4H-SiC). Silicon dioxide (\sioo) grown by plasma-enhanced chemical vapor deposition (PECVD) and by thermal oxidation of the \sioo-semiconductor interface are compared with respect to interface and defect formation. The nanometer depth resolution of \LEmuSR allows for a clear distinction between the oxide and semiconductor layers, while also quantifying the extension of structural changes caused by the oxidation of both Si and SiC. 

% \judith{(Needs revision!)}
\end{abstract}

% Do not abbreviate Figure, Equation, etc.; display items are always singular, i.e., Figure 1 and 2.
% Equations are always singular, i.e., Equation 1 and 2, and should be inserted using the {equation} environment, not as graphics
% Please do not use footnotes in the text, additional information can be added to the Reference list.

\section{\label{sec:intro}Introduction}

At present, silicon (Si) is still the most commonly used semiconductor for most applications, however, high expectations are raised by silicon carbide (4H-SiC), a wide-bandgap semiconductor exhibiting a high breakdown voltage and thermal conductivity, but also high chemical stability.\cite{Choyke_1997} 
Being a robust and biocompatible material that can sustain harsh environments, SiC is also used in the biomedical field.\cite{Saddow_2022} More recently, SiC proved to be a promising platform for quantum technologies, with implementation as single photon emitter for quantum communication and sensors.\cite{Son_2020,Bathen_2021} 
With respect to other semiconductors, a major advantage of Si and SiC is their ability to form native silicon dioxide (\sioo). Silicon dioxide is an excellent insulator with a large dielectric strength and high temperature stability. Apart from its favourable dielectric properties, it is also inert to most chemicals, can act as a diffusion barrier, and since it is easy to grow, etch and pattern, it plays a crucial role for most device manufacturing processes. For many applications, the growth of a high-quality oxide-semiconductor interface with low densities of defects is critical. Over the past decades, continuous efforts have been placed in optimizing oxidation processes for better device performance and reliability. However, despite the tremendous progress during the last decades in improving the quality of the \sioo/Si and \sioo/SiC interfaces, characterization and understanding of oxidation-induced defects - both in the oxide and the semiconductor - remain challenging.

Electrical properties of oxide-semiconductor interfaces are commonly obtained using capacitance-voltage (C-V) or conductance-voltage (G-V) measurements of metal-oxide-semiconductor (MOS) structures.\cite{Sze_2007,Schroder_2015} 
While these techniques allow for a very accurate determination of the energy position of oxide or interface defects, only limited information on the defect's nature or their spatial distribution with respect to the interface can be obtained.

Based on angle-resolved X-ray photoemission spectroscopy (AR-XPS), transmission electron microscopy (TEM) and electron-energy-loss-spectroscopy (EELS), it is commonly assumed that the majority of defects are mainly located in the narrow transition region between the semiconductor and the oxide and that there is a direct correlation between the width of this transition region and the performance of the MOS devices. However, there is a large spread of experimental data on the extension of this sub-stoi\-chiometric region, ranging from several \si{\angstrom} to tens of \si{\nano\meter},\cite{Kasap_Capper_2017, Hornetz_1994, Taillon_2013, Biggerstaff_2009} and an unambiguous interpretation of the structural information is not always given.

A powerful technique for the investigation of semiconductor defects is muon spin rotation spectroscopy ($\mu$SR),
\cite{muon_spectroscopy_2021,hillier_muon_2022} an atomic, molecular and condensed matter experimental technique based on nuclear detection methods. The technique usually uses a beam of spin-polarized, positively charged muons, unstable elementary particles with a lifetime of \SI{2.2}{\micro\second}. The muons are implanted into the target material where they act as a sensitive probe to their local electronic and magnetic environment before decaying into a positron and two neutrinos. The decay positron is anisotropically emitted, preferentially in the direction of the muon spin at the time of decay and is detected by an array of scintillators placed around the sample chamber. The decay asymmetry \textit{A(t)} is the signal obtained from the recorded decay positrons along the different detection directions, and determines the time evolution of the muon spin ensemble polarization.
%, therefore providing information about the evolution of the muon spin polarization in the host material. 
When the muon is implanted in insulators and semiconductors, it can thermalize as an unbound muon, or it can pick up one or even two electrons forming different muonium configurations (Mu$^+$, Mu$^0$ and Mu$^-$). Since the Mu states are very sensitive to interaction with charge carriers and defects, the formation probability of each Mu state is strongly dependent on its local surroundings.\cite{Percival_1979,Patterson_1988,Cox_2009,Eshchenko_2002,Alberto_2018,Prokscha_2020} In $\mu$SR, neutral and charged muonium states can be identified by the spin precession frequency in an applied magnetic field: \Mup, like a free muon, will have a spin precession equal to the muon's Larmor frequency, while \Muz has a faster precession (about 103 times in a low magnetic field of the order of mT) due to the strong hyperfine coupling between the muon and the electron. The other charged state, \Mum, is indistinguishable from \Mup by spectroscopic methods, because the hyperfine coupling of the muon with the two electrons with opposite spin cancels. The charged Mu configurations are called diamagnetic states, to distinguish them from the paramagnetic \Muz state.

A compelling extension to conventional bulk $\mu$SR is low-energy (LE) $\mu$SR which allows to study thin samples and multi-layered structures with a depth-resolution of a few nanometers.\cite{Morenzoni_2004} Low-energy muons are obtained by the moderation of typically \SI{4}{\mega\electronvolt} muons to almost thermal energies and subsequent acceleration by an electrostatic field.\cite{Harshman_1987,Morenzoni_1994,Morenzoni_2000,Prokscha_2001} By varying the energy in the range of \SIrange[range-units=single]{1}{25}{\kilo\electronvolt}, one can control the mean muon stopping depth. Depending on the material, it is possible to probe specific regions close to the surface and interfaces in a depth-resolved manner up to a depth of about \SI{200}{\nano\meter}.  \\
Recently, LE-$\mu$SR was successfully employed to study defects and band-bending effects near the surface in a variety of semiconductors such as Si, SiC, germanium (Ge), CdS or ZnO.\cite{Woerle_2019a,Woerle_2020,Prokscha_2013,Prokscha_2020,Prokscha_2014a} 
%
%In Si and Ge, muonium can form at the tetrahedral site (\MuT) and at the center of the S-S bond (\MuBC). \MuT has the most significant contribution from \textit{prompt} formation, meaning it is the result of the slowing down process occurring in the first few picoseconds upon implantation during charge-exchange processes. However, the formation of \MuBC depends on the availability of excess electrons, as seen by its increase with increasing muon implantation energy, that can be tuned to generate between a few to several thousand free electrons \cite{Eshchenko_2002,Prokscha_2007}. Thus, \MuBC is most likely to be the result of \textit{delayed} capture of an electron created in the muon ionization track. 
%
For example, investigation of the near-surface of p-type Ge enables the direct observation of charge-carrier profiles in the hole depletion region and the permanent removal of the depletion layer after illumination with a blue laser ($\lambda=$ \SI{457}{\nano \meter}).\cite{Prokscha_2020} As the sample is illuminated, photogenerated electrons occupy empty surface acceptor states and attract holes into the hole depletion layer.
\LEmuSR has also been successfully deployed to obtain the profile of defects in the near-surface region of SiC, where the suppression of \Muz formation is attributed to the presence of defects in the crystal.\cite{Woerle_2019a} In particular, the \muSR signal shows a distinct behavior for samples with large densities of either carbon or silicon vacancies.\cite{Woerle_2020} Here the interaction of the \mup with Si vacancies favors the formation of \Muz, while the C vacancy center prompts an electron double capture to form \Mum, resulting in an enhanced diamagnetic signal. 

%Polarized photoluminescence measurements of thermally oxidized 4H-SiC showed an alignment of the oxide defects along the main crystal axes, indicating a close proximity of the defects to the SiC bulk \cite{Lohrmann_2015}. 

In this study, we take advantage of the \LEmuSR depth-resolution to investigate interfacial systems, with focus on the muon probe response to either thermally grown or deposited \sio on both Si and 4H-SiC. The chosen set of samples intends to give an insight into the sensitivity of the technique to different parameters such as defect and charge carrier concentrations in the \sioo-semiconductor systems. Differences in both the oxide and the interface quality after the two oxidation treatments are observed, where the \LEmuSR data reveal a \SIrange[range-units=single]{10}{30}{\nano \meter} wide interface region with enhanced defect concentrations at the thermally grown \sio/Si and \sio/4H-SiC interfaces. In addition, in SiC, near the interface, the diamagnetic signal is sensitive to changes in charge carrier concentration, which depends on the initial doping concentration and the distinct oxidation processes of the Si- and C-face.

\section{\label{sec:results}Results and Discussion}
\begin{figure}[h]
	    \centering
	    \includegraphics[width=0.9\textwidth]{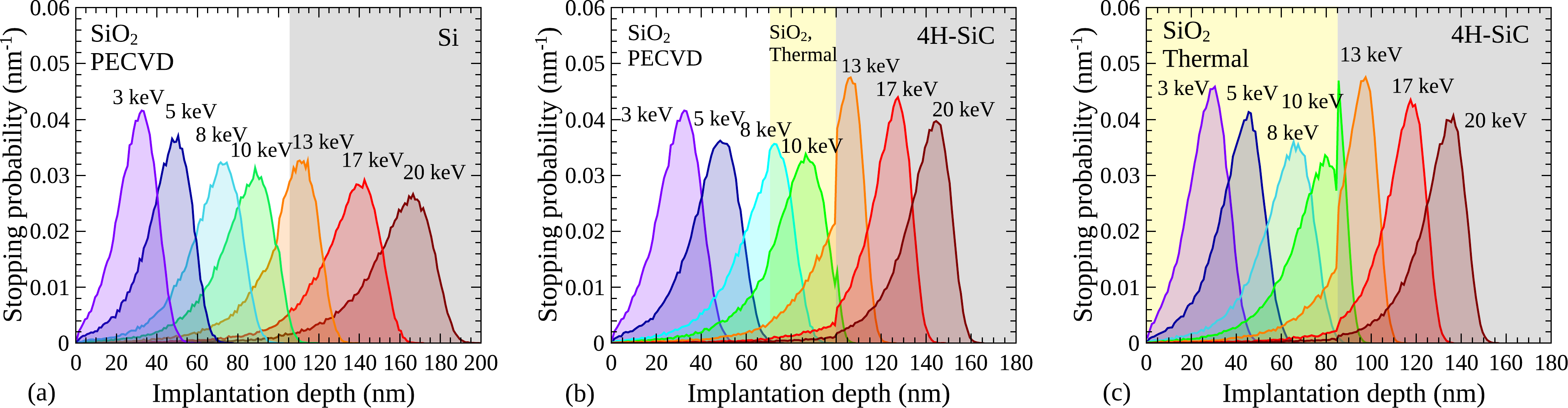}
	    \caption{Simulated stopping profiles for three of the studied oxide-semiconductor samples obtained with TRIMSP. (a) Si-A with a PECVD oxide of \SI{105}{\nano\meter}. (b) SiC-B with both a thin thermally grown and a deposited \sio on top of 4H-SiC. (c) SiC-C with a thermally grown \sio layer on top of 4H-SiC.}
	    \label{fig:Stopping_profiles}
	\end{figure}

An overview of the investigated samples for this \LEmuSR study is given in \Cref{tab:SiSamples,tab:SiCSamples}. A \sio film was formed on all samples, either by a low-temperature deposition in a PECVD chamber or by thermal growth in O$_2$ ambient. As the focus of this study was the investigation of interface defects and their effect on the \LEmuSR signal, no post-oxidation annealing treatment was performed for any of the samples. 
The stopping profiles of the muons are simulated for each sample using the TRIMSP code,\cite{Eckstein_1991,Morenzoni_2002} and provide the mean stopping depth of the muon beam for a given implantation energy. The stopping profiles are calculated based on the measured densities of the two oxides and are presented for three of the samples in \Cref{fig:Stopping_profiles}. For muon implantation energies $<$\SI{10}{\kilo\electronvolt}, the oxide layer is probed, while energies $>$\SI{16}{\kilo\electronvolt} mainly probe the semiconductor bulk. In order to study the interface and near-interface region, implantation energies between these two values are used.

In $\mu$SR, the so-called dia- and paramagnetic decay asymmetries \AD and \AMu are determined by the amplitudes of the muon spin precession signals in an applied magnetic field transverse to the initial muon spin direction (transverse-field TF-$\mu$SR). These asymmetries are proportional to the fraction of muons in the particular state. The respective fractions are calculated as: \FD =\AD/A$_{\mathrm{total}}$, and \FMu=2$\cdot$\AMu/A$_{\mathrm{total}}$, where A$_{\mathrm{total}}$ is the maximally observable decay asymmetry of the $\mu$SR spectrometer. The factor of 2 in \FMu results from the fact that only \SI{50}{\percent} of the total \Muz polarization is observable in our experiment with an applied magnetic field of \SI{0.5}{\milli\tesla}.\cite{muon_spectroscopy_2021}
If no \Muz is formed, the diamagnetic fraction \FD is one. Smaller values - normally observed in insulators and semiconductors - indicate the formation of \Muzo. At low temperatures and moderate doping of the semiconductor, more than \SI{90}{\percent} of the implanted \mup are expected to form such a paramagnetic state in Si or 4H-SiC.\cite{Kreitzman_1995,Woerle_2020} Very often, some fraction of \Muz states formed in insulators or semiconductors does not contribute to the precession signal due to fast muon spin depolarization processes, leading to a reduction of \AMu and to the observation, that, in general, \FD + \FMu $\le 1$ (which is called the \textit{missing fraction} of muon spin polarization in $\mu$SR). 

The depth dependence of \FDo($x$) and \FMuo($x$) was obtained from the correlation between implantation energy of the muon and mean stopping depth illustrated by the stopping profiles, with the fitting procedure detailed in \Cref{sec:experimental}. The simple model assumes \FD and \FMu to be constant within each layer, and changing abruptly at the interface of two layers. As will be discussed below, for the deposited \sioo samples, the fit result is consistent with the the measured thickness of the oxide, highlighting the accuracy of the analysis.
%this description.

\begin{table}[ht!]
\centering
     \caption{Si samples with donor concentration N$_D$ and oxide thickness t$_{Ox}$. All samples are cut from the same (100) Si wafer.}
 \begin{tabular}[htbp]{@{}lccccr@{}}
    \hline
    Name & Crystal Orientation & N$_{D}$ (\si{\per\centi\meter\cubed}) & Oxidation Process & t$_{Ox}$ (\si{\nano\meter}) \\
    \hline
    Si-A & (100)  & \SI{5e16}  & PECVD &  105 \\
    Si-B & (100) & \SI{5e16}  & thermal &  110 \\
    Si-C & (100)  & \SI{5e16}  & thermal, HF, PECVD & 95 \\
    \hline
 \end{tabular}
  \label{tab:SiSamples}
\end{table}

\subsection{\label{subsec:si_results}The SiO$_{2}$/Si Interface}

\begin{figure}[ht]
	    \centering
	    \includegraphics[width=0.85\textwidth]{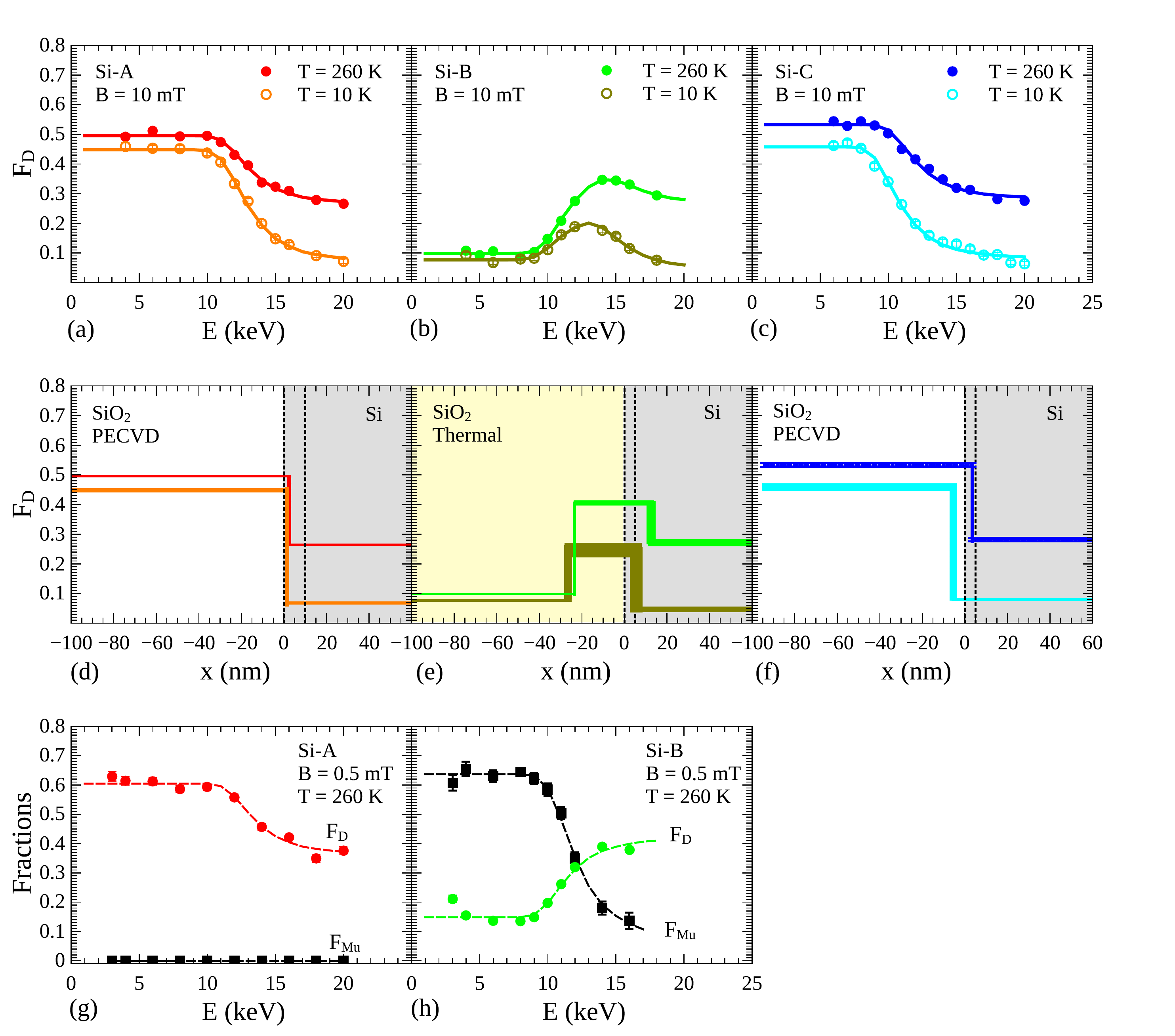}
	    \caption{Analysis of diamagnetic (\FD) and paramagnetic (\FMu) fractions measured in the Si samples. (a-c) \FD as function of muon implantation energy measured at \SIlist[list-units=single]{10;260}{\kelvin} with an externally applied magnetic field of \SI{10}{\milli\tesla}. (d-f) Depth variation of \FD obtained by fitting the corresponding muon implantation energy dependence. The width of the colored lines indicates the standard deviation of the fit parameters. (g-h) \FD and \FMu as function of muon implantation energy measured at \SI{0.5}{\milli\tesla}. The fitted dashed lines describe the variation of the fractions, assuming \FDo($x$) and \FMuo($x$) are abruptly changing between the different layers of the sample. }
	    \label{fig:Si_fits}
	\end{figure}

The results of the LE-$\mu$SR measurements (\FDo($E$)) at \SI{10}{\milli\tesla} for the \sioo/Si interface are summarized in \Cref{fig:Si_fits}~(a-c) and corresponding fit results \FDo($x$) are shown in \Cref{fig:Si_fits}~(d-f). The vertical dotted lines represent the uncertainty in determining the exact interface position by profilometer and X-ray reflectivity measurements. Additionally, for Si-A and Si-B, \FMu was extracted from the \SI{0.5}{\milli\tesla} measurements and are shown in \Cref{fig:Si_fits}~(g-h).

In \sioo, typically more than \SI{60}{\percent} of the muons form \Muz, and almost \SI{30}{\percent} of the implanted muons decay in the diamagnetic state.\cite{Kiefl_1982,Prokscha_2007,funamori_muonium_2015} In PECVD-grown \sioo (\Cref{fig:Si_fits}~(a)), the diamagnetic fraction is about \SI{50}{\percent}, possibly due to electron trapping in the low quality oxide, as hinted by the suppression of \Muz formation (\Cref{fig:Si_fits}~(g)).\cite{Prokscha_2014a,Woerle_2019a} In the case of the thermally grown oxide shown in Fig. \ref{fig:Si_fits}~(b), the diamagnetic fraction for low implantation energies is significantly lower than for the deposited oxide and comparable to what was previously reported for \sio crystal or glass.\cite{Prokscha_2007} Measurements performed with an externally applied magnetic field of \SI{0.5}{\milli\tesla}, where the \Muz procession signal can be directly observed, show a \FMu of \SI{65}{\percent} (\Cref{fig:Si_fits}~(h)), suggesting a higher structural order of \sio and hence an undisturbed \Muz formation process in the \sio bulk.

Another remarkable difference is the effect each oxidation process has on the interface formed with the semiconductor. The PECVD-\sio, deposited at low temperatures, can be clearly distinguished from Si in the \muSR signal, as \FD quickly drops to the Si layer value at the interface (\Cref{fig:Si_fits}~(d,f)). On the other hand, for sample Si-B with a thermally grown oxide, \FD increases around the \sioo/Si interface, suggesting a defect-rich region or a region with higher structural disorder mainly on the \sio side ($\sim$ \SI{20}{\nano\meter}), but also on the Si side (\SIrange[range-phrase={--},range-units=single]{5}{10}{\nano\meter}) of the interface. On the Si side, besides the presence of defects or structural disorder, the increase of \FD could be also generated by electron accumulation in the interface region.\cite{Prokscha_2020} 
In \sio, the existence of a defect-rich region is further supported by the \SI{0.5}{\milli\tesla} fitting results (in \Cref{fig:Si_fits}~(h)) where a conversion of \Muz to the diamagnetic state occurs at a muon implantation energy of \SI{12}{\kilo\electronvolt}, corresponding to a probing depth of ~\SI{20}{\nano\meter} away from the interface towards the oxide side. Thermal oxidation of Si is known to result in a large number of point-like defects in the \sio (E' centers) and dangling bonds (P$_{b}$ centers) at the interface.\cite{Schroder_2015,Kasap_Capper_2017}
These defects with energy levels in the bandgap can serve as traps for carriers and hence explain the observed differences of the diamagnetic signal between samples Si-A and Si-B. Furthermore, the growth of a thermal oxide on Si was previously shown to introduce stress into the interface system,\cite{Queeney_2000} resulting in a stress component both in the oxide, as well as in a narrow region in the Si.\cite{Kobeda_1987} Although the oxidation-induced strain is expected to quickly relax away from the interface, stress-related lattice distortion may contribute to the observed reduction of \FMu in \Cref{fig:Si_fits}~(h). However, in our samples, this strain-induced reduction of \FMu cannot be separated from the fast depolarization of \Muz in Si due to the presence of free electrons/dopants with 
N$_{D}$ = \SI{5e16}~\si{\per\centi\meter\cubed},\cite{Patterson_1988} as the detection of strain-induced effects on \FMu would require a Si sample with N$_{D} < 10^{13}$~\si{\per\centi\meter\cubed} to significantly reduce the dopant-induced depolarization of \Muz.

When removing the thermally grown \sio from the Si sample and depositing a PECVD-\sio instead (\Cref{fig:Si_fits}~(c)), the defective region around the interface disappears and a very similar \FD as for sample Si-A is observed. Interestingly, without the thermal oxide, the \FD on the Si side also recovers to the same values as for the deposited oxide, suggesting that the Si is not permanently altered by the thermal oxidation. 

At energies $\gtrsim$\SI{16}{\kilo\electronvolt} where the Si bulk is probed, all three samples show %comparable 
similar diamagnetic fractions of $\sim$\SI{10}{\percent} and $\sim$\SI{27}{\percent} for temperatures \textit{T} = \SIlist{10; 260}{\kelvin}, respectively, indicating that the Si remains unchanged from the oxidation process at these depths. 
%
% Thomas: the increase of diamagnetic fraction is mainly due to the ionization of MuBC above ~150 K.
% The donor concentration of 5e16 is probably not large enough to observe the increase of F_D due to
% Mu- formation. F_D should be > 30%, if Mu- formation takes place.
%
%
The larger value of \FD at \textit{T} = \SI{260}{\kelvin} is a result of the thermally activated ionization of one of the two \Muz states in Si: \Muz can either form at the bond-center between two Si atoms (\MuBC), or at the tetrahedral interstitial site (\MuT).\cite{Patterson_1988,Kreitzman_1995} It is the \MuBC state with a fraction of
$\sim$\SI{30}{\percent} which ionizes at \textit{T} $>$ \SI{150}{\kelvin} to form the diamagnetic Mu$_{BC}^+$ state \cite{Kreitzman_1995, fan_optically_2008,Prokscha_2012} that can be observed in TF-$\mu$SR 
at \SI{260}{\kelvin}.

	    \subsubsection{Electrical characterization of the \sio}
	    	\begin{figure}[ht!]
	    \centering
	    \includegraphics[width=1.0\textwidth]{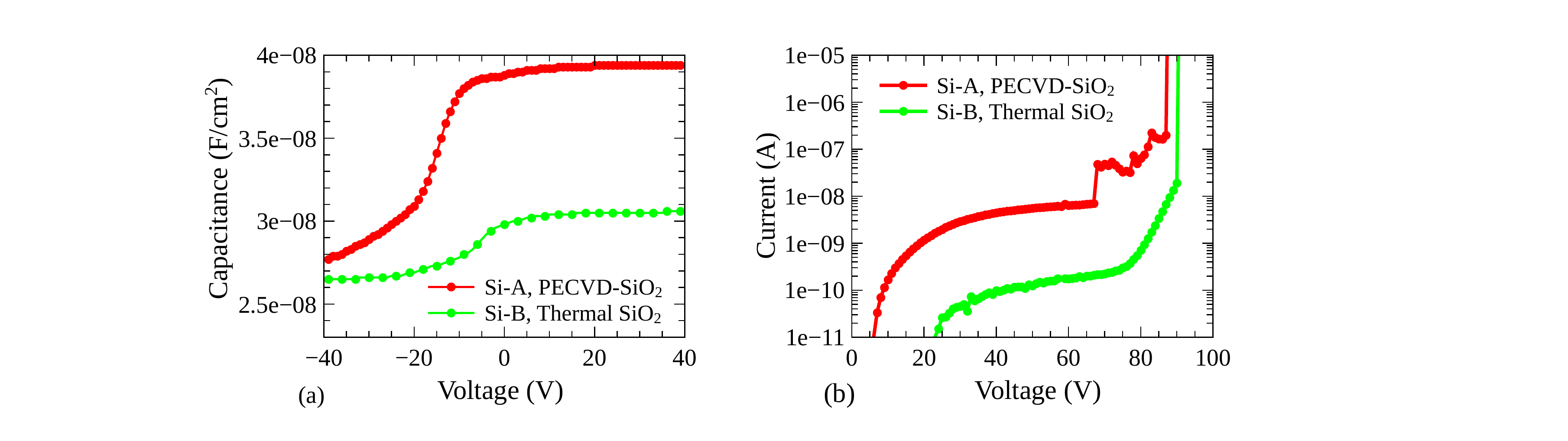}
	    \caption{Analysis of MOS capacitors fabricated on samples Si-A and Si-B. (a) Capacitance-voltage curves measured at \SI{1}{\mega\hertz}. (b) Current-voltage curves, showing an increased leakage current for the deposited \sioo film. All measurements were performed in the dark and at room temperature. The diameter of the circular contacts was \SI{400}{\micro\meter}.}
	    \label{fig:Si_electChar}
	\end{figure}
	    \sio layers formed by thermal oxidation or chemical vapor deposition are known to yield oxides with very different physical and electrical properties which is clearly reflected in the $\mu$SR measurements. In order to connect our microscopic analysis to macroscopic device properties, an electrical analysis of metal-oxide-semiconductor (MOS) capacitors fabricated on the Si samples was performed. Capacitance-voltage (C-V) and current-voltage (I-V) measurements of the MOS capacitors are presented in \Cref{fig:Si_electChar}.
	    
	    Despite the similar physical thickness of the two oxide layers, the C-V analysis reveals a much larger oxide capacitance for the deposited \sio (\SI{3.9e-8}{\farad\per\centi\meter\squared}) compared to the thermally grown film\\
	    (\SI{3.1e-8}{\farad\per\centi\meter\squared}). This behavior may be explained by a larger water content for the deposited \sio, causing an increase of the dielectric constant ($\epsilon_r$) from $\epsilon_r$ = \num{3.88} for the thermally grown oxide to $\epsilon_r$ = \num{4.67} for the PECVD-\sio.\cite{Pavelescu_1992,Idris_1998} Extraction of the interface defect density \dit,\cite{Nicollian_1982} reveals fairly large \dit values for the thermally grown oxide (\dit~=~\SI{2e12}{\per\centi\meter\squared\per\electronvolt}) and the deposited oxide\\ (\dit~=~\SI{9e11}{\per\centi\meter\squared\per\electronvolt}) which is expected for oxidation processes without any further annealing treatments. However, this also implies that the observed increase of \FD around the interface of the thermally grown oxide (Si-B) can not solely be explained by the presence of dangling bonds at the \sio/Si interface. Instead, one should consider a combination of factors, including oxide charges and interface traps in the \sio and at the interface, point defect and charge accumulation in the Si as well as strain-induced lattice distortions across the interface.

	    Differences between the two oxides are also observed in the breakdown measurements shown in \Cref{fig:Si_electChar}~(b), where the porous PECVD-\sio shows almost two orders of magnitude higher leakage currents and an earlier breakdown compared to the thermally grown \sio.

	\subsection{\label{subsec:sic_results}The SiO$_{2}$/4H-SiC Interface}
	
	\begin{table}[ht!]
\centering

 \caption{SiC samples with donor concentration N$_D$ and oxide thickness t$_{Ox}$. All samples are cut from the same 4H-SiC wafer with a low-doped epitaxial layer on the Si-face and the highly doped substrate on the C-face.}
 \begin{tabular}[htbp]{@{}lccccr@{}}
    \hline
    Name & Crystal Orientation & N$_{D}$ (\si{\per\centi\meter\cubed}) & Oxidation Process & t$_{Ox}$ (\si{\nano\meter})\\
    \hline
    SiC-A & (0001)Si  & \SI{8e15}  & PECVD &  105\\
    SiC-B & (0001)Si & \SI{8e15}  & thermal + PECVD & 30 + 70 \\
    SiC-C & (000$\overline{1}$)C & $\sim$~\SI{1e19}  & thermal &  90 \\
    SiC-D & (000$\overline{1}$)C  &$\sim$~\SI{1e19}  & thermal, HF, PECVD & 95 \\
    \hline
 \end{tabular}
  \label{tab:SiCSamples}
\end{table}

	Unlike for the Si samples where we mainly discussed the muon response to different \sio films, in the case of the SiC samples we also studied how the doping concentration of the semiconductor impacts the \FD and \FMu formation process across the interface. A summary of the processing parameters of the SiC samples is given in \Cref{tab:SiCSamples}. While the \sio/SiC interface was formed on the low-doped (0001) Si-face for samples SiC-A and SiC-B, measurements on samples SiC-C and SiC-D were performed on the highly doped (000$\overline{1}$) C-face. As the oxide growth rate at \SI{1050}{\celsius} for (0001) 4H-SiC is too low to achieve the targeted oxide thickness of \SI{100}{\nano\meter}, only a \SI{30}{\nano\meter}-thick oxide was thermally grown on sample SiC-B and a \SI{70}{\nano\meter}-thick PECVD-\sio was deposited on top. In order to investigate any permanent change of the SiC due to thermal oxidation,  sample SiC-D was first thermally oxidized before the oxide layer was removed again and a PECVD-\sio was deposited instead.

\begin{figure}[ht!]
	    \centering
	    \includegraphics[width=0.9\textwidth]{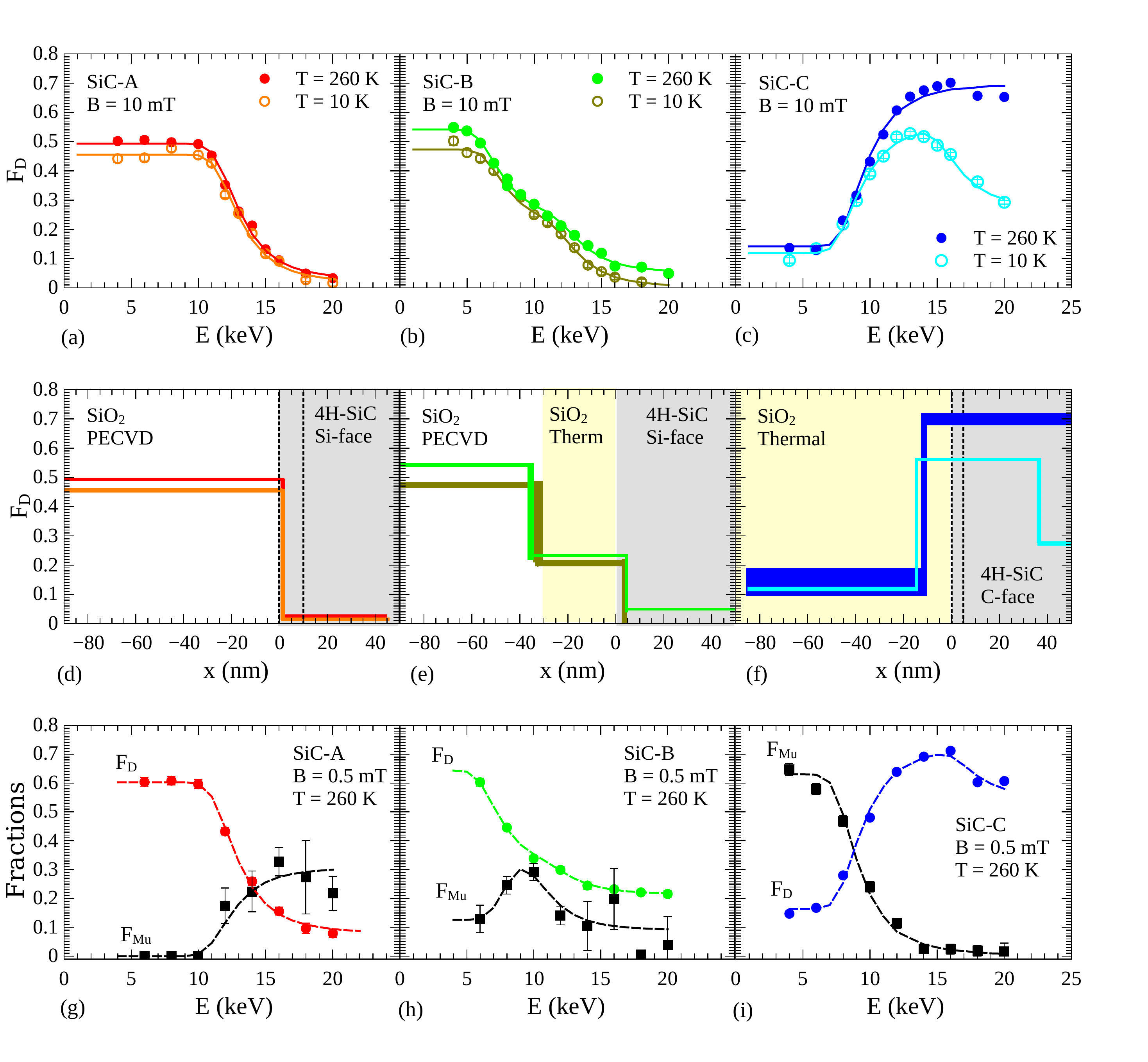}
	    \caption{Analysis of diamagnetic (\FD) and paramagnetic (\FMu) fractions measured for SiC-A, SiC-B and SiC-C. (a-c) \FD as a function of muon implantation energy measured at \SIlist[list-units=single]{10;260}{\kelvin} with an externally applied magnetic field of \SI{10}{\milli\tesla}. (d-f) Depth variation of \FD obtained by fitting the corresponding muon implantation energy dependence. The width of the colored lines indicates the standard deviation of the fit parameters. (g-i) \FD and \FMu as function of muon implantation energy measured at \SI{0.5}{\milli\tesla} and \SI{260}{\kelvin}. The fitted dashed lines describe the variation of the fractions, assuming \FDo($x$) and \FMuo($x$) are abruptly changing between the different layers of the sample.
	    }
	    
	    \label{fig:SiC_fits}
	\end{figure}

\begin{figure}[ht!]
	    \centering
	    \includegraphics[width=0.7\textwidth]{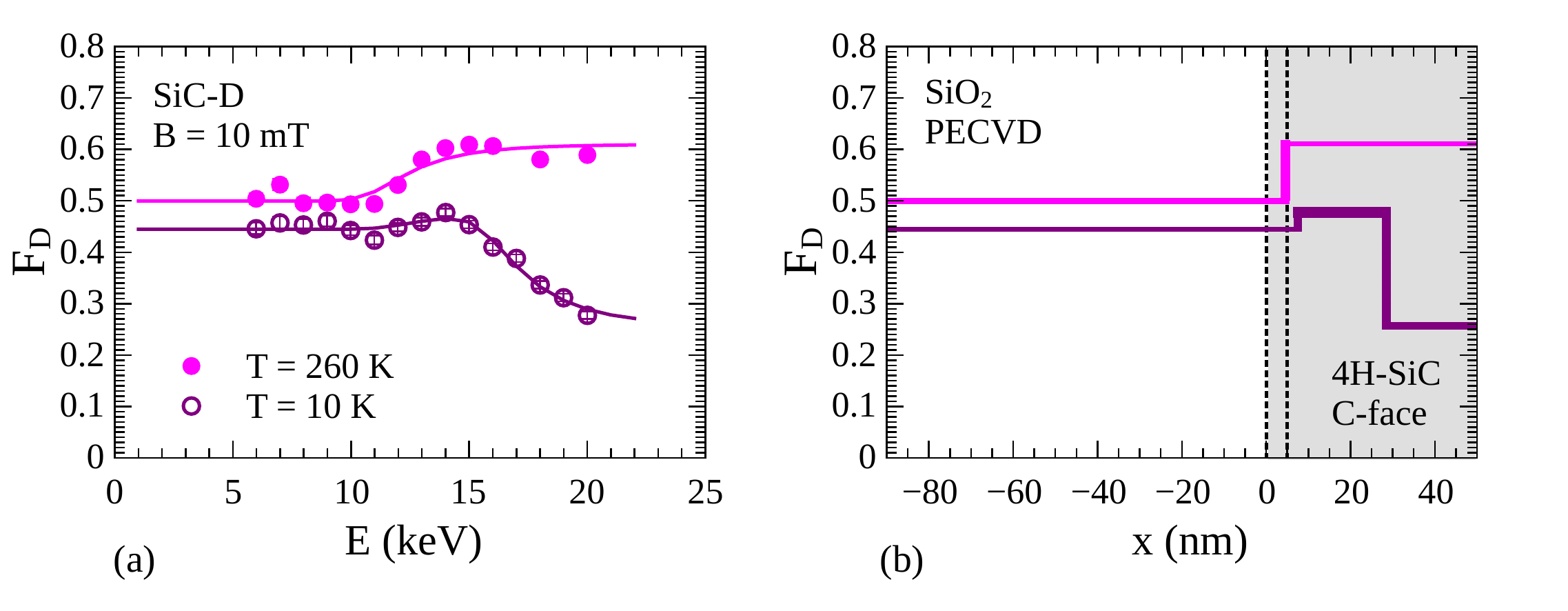}
	    \caption{ (a) Analysis of the diamagnetic fraction \FD measured in SiC-D,  as function of muon implantation energy at \SIlist[list-units=single]{10;260}{\kelvin} with an externally applied magnetic field of \SI{10}{\milli\tesla}. (b) Depth variation of \FD obtained by fitting the corresponding muon implantation energy dependence. The width of the colored lines indicates the standard deviation of the fit parameters.}
	    \label{fig:SiC_fits_SiCD}
	\end{figure}

	\Cref{fig:SiC_fits}~(a-c) and \ref{fig:SiC_fits_SiCD}~(a) show \FD measured on the 4H-SiC samples with an applied magnetic field of \SI{10}{\milli\tesla}, measurements performed at \SI{0.5}{\milli\tesla}  are presented in \Cref{fig:SiC_fits}~(g-i). Again, the vertical dotted lines represent the uncertainty in determining the exact interface position by profilometer and X-ray reflectivity measurements.%   We don't need a new paragraph here; therefore I commented this empty line. Thomas.
%	We don't need a new paragraph here; therefore I commented this empty line. Thomas.
As shown before for the Si samples, the depth-dependence of \FDo($x$) in \Cref{fig:SiC_fits}~(d-e) clearly indicates the \sioo/SiC interface where \FD quickly drops to the typical 4H-SiC bulk value.\cite{Woerle_2019a,Woerle_2020}

%comparison of oxides
	The PECVD oxide of samples SiC-A, SiC-B, and SiC-D shows a similar behavior as on Si with a strong increase of \FD compared to the bulk SiC and a suppression of \Muz formation across the whole oxide layer.  
	Similarly to Si-B, the thermal oxide in the SiC-B and SiC-C samples has a higher structural order compared to PECVD-\sioo, allowing observation of the \Muz precession in the \sioo layers. %This is also supported by the \dit analysis, indeed revealing a larger interface defect density (\dit~=~\SI{8e12}{\per\centi\meter\squared\per\electronvolt}) for the thermally grown oxide (SiC-B) compared to the Si samples or the deposited \sio on sample SiC-A (\dit~=~\SI{1e12}{\per\centi\meter\squared\per\electronvolt}).
	
	Thermal oxidation of Si and SiC is very similar, however, in the case of SiC, it does not only involve the formation of bonds between Si and O, but also removal of C through the oxide. If the carbon atoms are not efficiently removed from the interface, they may form carbon-related defects directly at, or in the vicinity of the \sio/SiC interface.
	Hence, apart from dangling bonds, interfacial carbon clusters and near-interface oxide traps are expected to contribute to the interface defect state density.\cite{Dhar_2009} 
	%Besides, 
	Additionally, electrically detected magnetic resonance (EDMR) measurements also suggest the out-diffusion of Si from the SiC bulk and the generation of silicon vacancies as a source for the increased defect concentrations close to the \sio/SiC interface.\cite{Cochrane_2011, Anders_2015} The Si-out diffusion during thermal oxidation, leads to a C-rich surface where large areas of graphene-like mono-layers are likely to form. \cite{Hass_2008} 
	For the SiC samples discussed here, the \dit analysis revealed a larger interface defect density (\dit~=~\SI{8e12}{\per\centi\meter\squared\per\electronvolt}) for the thermally grown oxide (SiC-B) compared to the Si samples or the deposited \sio on sample SiC-A (\dit~=~\SI{1e12}{\per\centi\meter\squared\per\electronvolt}).
	The effect of enhanced defect concentrations in the thermally grown \sio layer correlates with the observed increase of \FD in the narrow \SI{30}{\nano\meter} thick thermal-\sioo layer in SiC-B.
	%Furthermore, the slower thermal oxidation of the Si-face of SiC compared to Si 
	%means it was only possible 
	%allowed to grow only a narrow \SI{30}{\nano\meter} thermal-\sioo layer in SiC-B.
	Although \Muz is visible in the thermal-\sio layer of SiC-B (\cref{fig:SiC_fits}~(h)), the enhanced \FD of \SI{22}{\percent} in \Cref{fig:SiC_fits}~(e) is comparable to the value measured at \SI{10}{\kelvin} in the defective \sio region near the interface of Si-B (\Cref{fig:Si_fits}~(e)). Thus, in SiC-B the \SI{30}{\nano\meter} layer of thermally grown \sioo seems to be affected by defects and oxidation-induced stress, which is reported to be more pronounced across the \sioo/SiC interface than that of \sioo/Si,\cite{Yoshikawa_2019} and also appears broader than the oxide defective region in Si-B and SiC-C.

%	\subsubsection{Impact of Crystal Orientation}

	 The quality of the oxide near the interface formed with SiC also depends on the polarity of the oxidized SiC face. The thermal oxidation process is almost ten times faster for the C-face of SiC than for the Si-face.\cite{Song_2004} Therefore, the resulting  oxide thickness using the same process parameters as for SiC-A and SiC-B results in a \SI{90}{\nano\meter} thermal-\sioo layer on the C-face in SiC-C. A region with enhanced \FD formation, like in Si-B, appears near the interface. The oxide defective region in SiC-C is narrower than in SiC-B, and affects only \SIrange[range-phrase={--},range-units=single]{10}{15}{\nano\meter} (\Cref{fig:SiC_fits}~(f)), which suggests less induced strain for \sio layers on the C-face.\cite{Hirai_2017} 
	 However, the higher \FD for SiC-C, compared to Si-B and SiC-B, suggests the presence of a much higher \dit for \sioo grown on C-face.\cite{Afanasev_1997,Fukuda_2000}
	 %as confirmed by the \dit measurements in the previous section. 
	 % Thomas:
	 % I was confused: the \dit measurements are for Si-A/B and SiC-A(B only, we don't have it for SiC-C, right? 
	 %%Maria:
	 % Exactly- it was not possible to measure dit for SiC-C/D due to the high doping of the C-face SiC.
	 %
	 The extension of the region with enhanced \FD formation into SiC also differs between Si- and C-face samples: it is $\sim$ \SI{5}{\nano\meter} in SiC-B, much narrower compared to the $\sim$
	 \SI{30}{\nano\meter} in SiC-C and SiC-D. The graphitization of SiC is considered to strongly contribute to the observations since in Si-face SiC only mono-layers of graphene are expected to form at \SI{1250}{\celsius}, whereas the C-face graphitizes at lower temperatures and generates three-dimensional layers. \cite{Luxmi_2010, Srivastava_2012} Thus, the \SI{30}{\nano\meter} wide region with a sharp rise in \FD observed in C-face SiC-C and SiC-D is likely related to the formation of a graphene-like layer in SiC near the interface.
	 However, this difference cannot be solely attributed to the impact of crystal orientation on the oxidation process, as it is indistinguishable from the effect of different doping concentrations on \FD at the surfaces of the semiconductors, where SiC-B has a three orders of magnitude smaller doping concentration than SiC-C. 
	 %(SiC-B: N$_D$ = \SI{8e15}~$\si{\per\centi\meter\cubed}$,
	 %SiC-C: N$_D$ = \SI{1e19}~$\si{\per\centi\meter\cubed}$).
%	 have on the surface of the semiconductor.
	
% \subsubsection{Formation of an accumulation region}

In SiC-A, \FD is nearly temperature-independent (\Cref{fig:SiC_fits}~(a,d)) in the SiC crystal where the \Muz signal is visible, and the respective fractions in \Cref{fig:SiC_fits}~(g) are comparable to similarly doped epitaxial SiC reported by Woerle et al.\cite{Woerle_2019a,Woerle_2020}
Although the doping concentration of SiC-B is the same as of SiC-A, the thermal oxidation in SiC-B leads to an unexpected increase of \FD to $\sim$~20\% within the semiconductor at \SI{0.5}{\milli\tesla} (\Cref{fig:SiC_fits}~(h)). Defects related to the carbon vacancy in SiC have been shown to lead to an increased \FD, while a paramagnetic fraction of $\sim$\SI{30}{\percent} still remains.\cite{Woerle_2019a} However, it is unlikely that carbon vacancies are responsible for the observations in \Cref{fig:SiC_fits}~(h), since the oxidation process itself is initiated by a Si out-diffusion and only a later destruction of the C sub-lattice.\cite{Woerle_2019} Furthermore, \LEmuSR is only sensitive to carbon defect concentrations $>$ \SI{1e17}{\per\centi\meter\cubed}, where \Muz still can be observed. In contrast, the \Muz signal disappears in the SiC layer of the SiC-B sample. Such a disappearance of the \Muz signal can be
explained by the interaction with free charge carriers, and formation of a diamagnetic state. The increase of \FD is hardly observable at \SI{10}{\milli\tesla}, which suggests that there is a neutral \Muz precursor state which quickly dephases.\cite{Woerle_2019a} At \SI{0.5}{\milli\tesla}, due to the smaller precession frequencies of the muon spin in \Muz compared to  \SI{10}{\milli\tesla}, the contribution of either electron or hole capture by \Muz to form \Mum or \Mup, respectively, can be observed. An accumulation of holes near the SiC surface would be in agreement with the reports of energy band structure of \sioo/SiC, which promotes depletion of electrons at the SiC interface.\cite{Watanabe_2011} Therefore it is likely that the \FD increase at \SI{260}{\kelvin} is due to delayed formation of \Mup, favoured by the band-bending caused by thermal oxidation of the Si-face SiC.

 %
 % Thomas:
 %
 % My memory about the data in the Patterson review was wrong. There are no FD data presented for Si and Ge, where F_D at high doping levels of n- or p-type can be compared. I would rewrite the section, as I suggest below. I think we should not mention the simulation and the formation rates - it doesn't seem to fit in the scope of this paper. Better, to present it in the paper about the different doping levels.
%

The high value of \FD $\sim$ \SI{55}{\percent} at \SI{10}{\kelvin} in a 30-nm-wide region inside the semiconductor in highly-doped, C-face SiC-C and SiC-D hints towards an electron-rich region near the interface.
The graphene layer grown on SiC has been reported to be n-type. \cite{Kopylov_2010} The doping is proposed to originate from three different mechanisms, which at times coexist: electron transfer from interface states (Si and C dangling bonds) to the graphene layer, polarization induced by the hexagonal geometry of the 4H-SiC substrate, and the effect of a space-charge region in doped SiC. Additionally, electron exchange with the bulk is also compatible with the C-face SiC substrate doping concentration of \SI{1e19}{\per\cubic\centi\meter}.  \cite{Ristein_2012, Mammadov_2014, Pradeepkumar_2020}
At a distance beyond \SI{30}{\nano\meter} from the interface, \FD drops to 
$\sim$ ~\SI{25}{\percent}. It is obvious to assume that the formation of this diamagnetic fraction in the "bulk" of the SiC originates from the capture of majority charge carriers (electrons) of \Muz. Comparing with unpublished data where we measured \FD for various n- and p-doping levels up to \SI{1e18}{\per\centi\meter\cubed}, a fraction of \FD~$\sim$ ~\SI{25}{\percent} corresponds to an electron concentration of $n \sim$~\SI{5e17}{\per\centi\meter\cubed}. 
%
% Thomas: here I used our data, that at 1e17, 10 mT, 260 K, we have FD ~ 15%, and at 1e18 FD ~ 35%.
%
This is much smaller than the bulk doping level of \SI{1e19}{\per\centi\meter\cubed}, since at 10~K, only a small fraction of donors is ionized. The increase of \FD to $\sim$ \SI{55}{\percent} closer to the interface would then mean an increase of $n$ to $\sim$~\SI{5e18}{\per\centi\meter\cubed} in the 30-nm-wide region of C-face SiC. %The out-diffusion of Si during thermal oxidation of C-face SiC leads to Si vacancies, which can then be occupied by nitrogen dopant atoms.
 Although an increase of \FD could also be sustained by hole capture of \Muz to form \Mup, the minority charge carrier concentration required to obtain \FD~=~\SI{55}{\percent} would have to be about five times larger than the $n$ estimated above, because the hole capture rate of \Muz is governed by the hole mobility, \cite{Prokscha_2014} which is about five times lower in 4H-SiC than the electron mobility. This would require a hole carrier concentration $>$~\SI{1e19}{\per\centi\meter\cubed}, which is larger than the n-type doping concentration.
%
% Additionally, the formation rate of \Mup and \Mum can be estimated with a Monte Carlo simulation of the TF-$\mu$SR histograms. \cite{Prokscha_2012,Prokscha_2014} The "bulk" \FD of SiC-C and SiC-D at \SI{10}{\kelvin} was used to estimate the charge carrier concentration to be $\sim$~\SI{8e17}{\per\centi\meter\cubed}. With such an electron concentration the \Mum formation rate obtained is \SI{800}{\mega\hertz}. The simulation results also show that for the case of \Mup formation at the same rate requires 5-7 times higher concentration of hole carriers. A higher concentration of holes compared to electrons is also reported in ref. \cite{Patterson_1988} for semiconductors Si and Ge, in order to observe a change in the charge state of muonium.  This can be explained by the low hole mobility of holes in SiC, compared to electrons, making the hole capture process a slower process.
 %
 %
 %
 Therefore, the remarkable increase of \FD in the highly doped C-face SiC, at the expense of \Muz (\Cref{fig:SiC_fits}~(f)),
 can be attributed to an enhanced concentration of electrons, in the n-type layer at the surface of SiC, leading to an increased \Mum formation probability. 
 %in that particular near-interface region or within the remaining C sublattice. 
 %Considering the previously discussed Si out-diffusion during thermal oxidation, a high density of Si vacancies ($\sim$~\SI{5e18}{\per\centi\meter\cubed}) seems to be generated in the 30-nm-wide region near the interface with C-face SiC. This together with the high concentration of dopant atoms ($\sim$~\SI{1e19}{\per\centi\meter\cubed}) can contribute to the observed increase in \Mum formation, if the Si vacancy is formed in close proximity to the N dopant (which is very likely, as the difference in concentration between dopant atom and vacancy is less than one order of magnitude).
 %
 Considering the previously discussed defect formation mechanisms during thermal oxidation, the change in carrier concentration can be attributed to the negatively charged graphene-like layer near the interface of the oxidized C-face.

 An interesting memory effect is observed in SiC-D at \SI{10}{\kelvin} (\Cref{fig:SiC_fits_SiCD}), where the thermal oxide was replaced by a deposited oxide. A region with high \FD is still present in C-face SiC even after removing the thermally grown \sio, suggesting that the oxidation process caused a permanent modification of the SiC crystal, e.g. by emission of oxidation by-products into the SiC crystal. Interestingly, \FD only increases a few nanometers below the \sio/SiC interface, suggesting a narrow C-rich layer, as discussed for SiC-C, which was not removed during HF etching,\cite{Johnson_2019} and is still present below the PECVD oxide.

In SiC-C and SiC-D, conversion of \FMu to \FD is visible in the semiconductor at \SI{260}{\kelvin} ((\Cref{fig:SiC_fits}~(i)), as the thermal ionization of nitrogen donors takes place at \textit{T} $>$ \SI{75}{\kelvin}, where the availability of free electrons from the ionized donors promote the formation of \Mum from \Muz precursor states. Thus, at \SI{260}{\kelvin} the n-type C-rich region where \FD is enhanced is no longer distinguishable due to the high concentration of free electrons. This enhanced \Mum formation at
260 K is the cause for the higher value of \FD $\sim$~70\% deep ($>$~40~nm) inside the semiconductor, compared to $\sim$~25\% at 10 K.

\section{\label{sec:conclusion}Conclusion}
In this study, we have demonstrated our recent progress on low-energy muon spin rotation for the investigation of oxide-semiconductor interfaces. LE-$\mu$SR offers a depth-resolved analysis of interfacial systems without the need for dedicated device structures or sophisticated system models for the interpretation of the data. The muon's sensitivity to defects and charge carriers allowed to distinguish characteristic properties of PECVD \sioo and thermally grown \sioo and their effect on the semiconductors: the high-temperature oxidation of Si and SiC results in an oxide with a superior structural order where the atom-like \Muz is likely to form, whereas in a PECVD-\sioo deposited at \SI{300}{\celsius} only a large \FD is observable. The samples with PECVD-\sioo exhibit an abrupt transition from the oxide to the semiconductor, whereas samples with thermally grown \sioo exhibit a \SIrange[range-phrase={--},range-units=single]{10}{30}{\nano \meter} wide region with enhanced \FD in the oxide near the interface. In the case of Si, the region with increased \FD at the interface inside the semiconductor disappeared once the thermal oxide was etched back and a PECVD-\sioo layer was deposited instead. In SiC, both the width of the region with enhanced \FD, and the strength of the diamagnetic signal induced by the thermal oxidation were shown to depend on the crystal orientation. In the C-face SiC samples, the region extends into the semiconductor and is still visible after etching the thermal oxide and deposition of PECVD-\sioo. The increase in \FD on the \sioo side of the interface is attributed to a combination of factors such as interface defects, near-interface oxide traps, and oxidation-induced stress. On the SiC side of the interface, the higher \FD in the first tens of nanometers is a result of the sensitivity of the \muSR signal to a change in charge carrier concentration at the SiC surface. These changes depend on the corresponding bulk carrier concentration and the overall oxidation process: on the Si-face depletion of electrons and hole capture sustains \Mup formation, and on the C-face, \Mum forms in the C-rich n-type layer. 

These results demonstrate that LE-$\mu$SR has the potential to deliver information about structural and electronic properties of oxide-semiconductor interfaces with hitherto inaccessible sensitivity and depth resolution. A good understanding of interface properties is critical for device operation, and the information provided by \LEmuSR can, in turn, advance the development of reliable SiC power devices.

% Experimental section

\section{Experimental Section}
\label{sec:experimental}
\threesubsection{Sample preparation}\\

For this experiment, (100) silicon with a nitrogen donor concentration of $N_D$ = \SI{5E16}{\per\centi\meter\cubed} as well as (0001) Si-face 4H-SiC ($N_D$ = \SI{8E15}{\per\centi\meter\cubed}) and (000$\overline{1}$) C-face 4H-SiC ($N_D \approx$ \SI{1E19}{\per\centi\meter\cubed}) were used. All samples were cut into sizes of \SI{25}{\milli\meter}\,$\times$\,\SI{25}{\milli\meter} and wet-chemically cleaned prior to oxidation. 
The oxide was either thermally grown at \SI{1050}{\celsius} in O$_2$ ambient or deposited at \SI{300}{\celsius} in a PECVD chamber. No post-oxidation annealing or other processes for improving the oxide quality were performed for any of the samples. The final oxide thicknesses were confirmed by profilometer and reflectometer measurements. Two samples, Si-C and SiC-D, were first thermally oxidized before the \sio was removed again by dipping the sample in hydrofluoric acid (HF) and another \sio layer was deposited on the samples. As the oxide growth rate at \SI{1050}{\celsius} is very low on the (0001) Si-face of 4H-SiC, only a \SI{30}{\nano\meter}-thick oxide was thermally grown on sample SiC-B and another \sio layer of \SI{70}{\nano\meter} was deposited on top.

\threesubsection{Density measurements}\\
Additional X-ray reflectivity (XRR) measurements proved helpful for the estimation of oxide densities: for thermally grown oxides, a density of $\rho_{\mathrm{SiO2, dry}}$ = \SI{2.2}{\gram\per\centi\meter\cubed} was extracted, whereas the deposited oxide had a slightly smaller density of $\rho_{\mathrm{SiO2, dep}}$ = \SI{2.1}{\gram\per\centi\meter\cubed}.

\threesubsection{Electrical characterization}\\
After the $\mu$SR experiment, MOS structures were fabricated on the samples and C-V and current-voltage (I-V) measurements were performed. The \dit values were determined from the capacitance and conductance curves obtained at \SI{1}{\mega\hertz}.

\threesubsection{$\mu$SR measurements}\\
The \LEmuSR experiments were performed at the low-energy muon facility (LEM) located at the $\mu$E4 beamline\cite{Prokscha_2008} of the Swiss Muon Source (S$\mu$S, Paul Scherrer Institute, Villigen, Switzerland). 
The samples were glued with conductive silver paint onto a Ni-coated aluminum sample plate, mounted  on a cold-finger cryostat. 
 The muon implantation energy ranged from \SIrange[range-units=single]{4}{20}{\kilo\electronvolt}, probing the first \SIrange[range-units=single]{150}{190}{\nano\meter} of the SiC and Si samples, respectively. The energy dependent measurements were carried out at temperatures of \SIlist{10; 260}{\kelvin}, with transverse magnetic fields of \SIlist[list-units=single]{0.5;10}{\milli\tesla} applied parallel to the beam axis. At a magnetic field of \SI{10}{\milli\tesla} only the diamagnetic signal can be resolved due to the frequency detection limit of \SI{50}{\mega\hertz} of the instrument. Additionally, at \SI{0.5}{\milli\tesla} the precession signal of \Muz is detected, where only about \SI{50}{\percent} of the total polarization is visible.\cite{muon_spectroscopy_2021} For \SI{10}{\milli\tesla} measurements 3 million low-energy muon events were recorded, and 8 million events for \SI{0.5}{\milli\tesla} measurements.

 The $\mu$SR asymmetry spectra were analyzed using the software \textit{musrfit}.\cite{Suter_2012}
 At a field of  \SI{10}{\milli\tesla}, where in our case only the diamagnetic signal 
 with asymmetry $A_\text{D}$ is observable, an exponentially-damped cosine function
 at the Larmor frequency $\omega_\mu$ of the muon was used to fit the data
 \begin{equation}
     A(t)=A_\text{D} \cdot \exp(-\lambda_\text{D} \cdot t) \cdot \cos(\omega_\mu t + \phi),
     \label{eq:Lorentzian1C}
 \end{equation}
where $\lambda_\text{D}$ is the depolarization rate of the diamagnetic state, and $\phi$ is the phase of the precession signal in a specific positron detector.
 At \SI{0.5}{\milli\tesla}, a second component becomes visible, which is the precession of the
 paramagnetic \Muz state with asymmetry $A_\text{Mu}$:
 \begin{equation}
     A(t)=A_\text{D} \cdot \exp(-\lambda_\text{D} \cdot t) \cdot \cos(\omega t + \phi)+ A_\text{Mu} \cdot \exp(-\lambda_\text{Mu}\cdot t)\cdot \cos(\omega_\text{Mu} t + \phi).
     \label{eq:Lorentzian2C}
 \end{equation}
 where $\lambda_\text{Mu}$ is the exponential depolarization rate of \Muz, $\omega_\text{Mu}$ is
 the muon spin precession frequency in the \Muz triplet state,\cite{muon_spectroscopy_2021} 
 and it is assumed that the diamagnetic and paramagnetic components have the same detector phase $\phi$.

\begin{figure}[h!]
	    \centering
	    \includegraphics[width=1\textwidth]{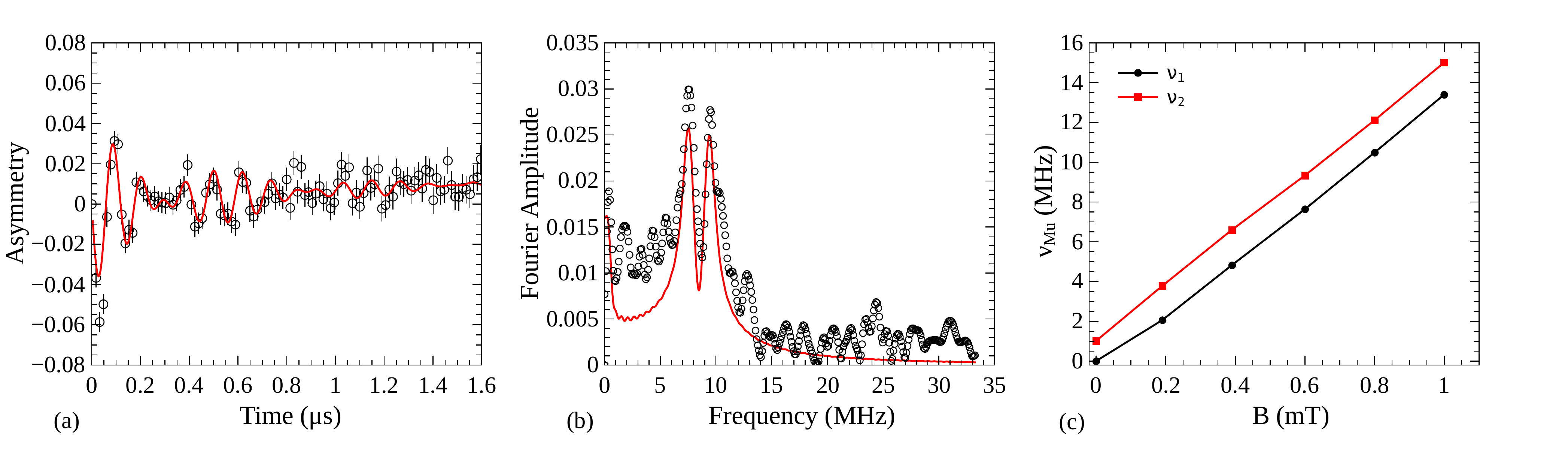}%
	    
	    \caption{ (a) \Muz asymmetry spectrum in a low-doped 4H-SiC epitaxial layer in an applied field of \SI{0.6}{\milli\tesla}. (b) Power Fourier spectrum of (a). (c) \Muz precession frequencies $\nu_{1(2)} = \omega_{1(2)}/2\pi$ as a function of applied field. $\Delta\nu = \nu_2 - \nu_1$ 
	    does not change in this field range, which means that $\nu_a = \Delta\nu \ll \nu_{iso}.$}
	    \label{fig:SiC_Muonium}
	\end{figure}

In 4H-SiC, the \Muz precession frequency splits into two lines due to a so far unreported weak anisotropy of the hyperfine coupling of $\nu_{a} \sim$~\SI{1.7}{\mega\hertz} on top of the isotropic part with a much larger coupling $\nu_{iso}$ of about \SI{3000}{\mega\hertz}:
 \begin{equation}
     A(t)=A_\text{D} \cdot \exp(-\lambda_\text{D} \cdot t) \cdot \cos(\omega t + \phi)+ A_\text{Mu} \cdot \exp(-\lambda_\text{Mu}\cdot t)\cdot [\cos(\omega_1 t + \phi) + \cos(\omega_2 t + \phi)],
     \label{eq:Lorentzian3C}
 \end{equation}
where $\omega_1 = \omega_\text{Mu}-2\pi\nu_{a}/2$, and $\omega_2 = \omega_\text{Mu}+2\pi\nu_{a}/2$, 
see Fig.~\ref{fig:SiC_Muonium}. The isotropic coupling $\nu_{iso}$ has been determined in Ref.~\cite{Lichti_2004}, where the authors indicated the presence of a presumably small anisotropic component in the hyperfine coupling. In the analysis of our data, the splitting of the \Muz lines in 4H-SiC allows to distinguish the fractions of muonium forming in the oxide and in SiC.

\threesubsection{$\mu$SR fitting procedure}\\
The method implemented by Sim{\~o}es et al. takes advantage of the nanometer depth-resolution of the \LEmuSR technique to infer the depth variation of the parameters from the experimentally measured energy dependence.\cite{Simoes_2020}
In this work, the depth dependence of the diamagnetic fraction is obtained using the correlation of the muon implantation energy and its stopping depth, the stopping probability $P$($x$,$E$), calculated via the Monte Carlo simulation TRIMSP \cite{Eckstein_1991,Morenzoni_2002} for all the samples. $P$($x$,$E$) is the probability per unit length that a muon implanted with energy $E$ stops in the material at a depth $x$, as shown for different samples in \Cref{fig:Stopping_profiles}. 

The fraction \FD of muons with a final diamagnetic state depends on the material, and has distinct characteristic values for the oxide and the semiconductor. Thus, we assume a step-like function to be a good approximation of the \FD variation within the structure.\cite{Alberto_2018,Curado_2020}

\medskip
%\textbf{Supporting Information} \par %Please delete the Suppporting Information statement if it is not applicable. Please supply Supporting Information in another file. Supporting information should not be provided in .tex format
%Supporting Information is available from the Wiley Online Library or from the author.

% Acknowledgements
\medskip
\textbf{Acknowledgements} \par %delete if not applicable))
The muon measurements have been performed at the Swiss Muon Source S$\mu$S, Paul Scherrer Institute, Villigen, Switzerland. We thank Dr. Laura Maurel Vel\`{a}zquez for performing the XRR measurements. 

% References
\medskip

\bibliographystyle{MSP}
\bibliography{library}

\end{document}